\documentclass{smarthepnote}
\usepackage[colorinlistoftodos]{todonotes}
\usepackage{titlesec}
\usepackage{placeins}
\usepackage{siunitx}
\usepackage{float}

\title{Review of Machine Learning for Real-Time Analysis at the Large Hadron Collider experiments ALICE, ATLAS, CMS and LHCb}
\author{The SMARTHEP Network}
\date{\today}

\usepackage{enumitem,amssymb}
\usepackage{pifont}

\DocAuthors{The SMARTHEP network}
\DocEditors{ L.~Boggia \\ C.~Cocha \\  F.~I.~Giasemis \\ J.~Hansen \\ P.~Inkaew \\ K.~E.~Iversen* \\ P.~Jawahar \\ H.~Pineiro Monteagudo \\ M.~Olocco \\ S.~Astrand*  }
\DocCoordinators{M.~Borsato\\ L.~Bozianu \\ S.~Schramm }
\DeliverableNo{D4.1}
\draftversion{2.0}

\begin{document}
\maketitle

\begin{abstract}
\vspace{-0.25cm}
The field of high energy physics (HEP) has seen a marked increase in the use of machine learning (ML) techniques in recent years. The proliferation of applications has revolutionised many aspects of the data processing pipeline at collider experiments including the Large Hadron Collider (LHC). In this whitepaper, we discuss the increasingly crucial role that ML plays in real-time analysis (RTA) at the LHC, namely in the context of the unique challenges posed by the trigger systems of the large LHC experiments. We describe a small selection of the ML applications in use at the large LHC experiments to demonstrate the breadth of use-cases. We continue by emphasising the importance of collaboration and engagement between the HEP community and industry, highlighting commonalities and synergies between the two. The mutual benefits are showcased in several interdisciplinary examples of RTA from industrial contexts. This whitepaper, compiled by the SMARTHEP network, does not provide an exhaustive review of ML at the LHC but rather offers a high-level overview of specific real-time use cases.
\vspace{0.25cm}
\end{abstract}

\vfill
\makereviewtable
\clearpage

\begingroup
\color{black}
\pagebreak

\section{Introduction}

The extraordinarily large datasets collected by the four Large Hadron Collider (LHC) experiments — ALICE, ATLAS, CMS and LHCb — since the start of LHC data-taking in 2009 represents the most precisely curated set of high-energy hadronic collisions ever recorded, but also an unparalleled technological challenge.
Collisions at the LHC occur at a frequency of up to $30$\,MHz, producing data at a rate of one petabyte per second in the current LHC Run 3 data taking period. 
Managing this huge data rate requires real-time processing to determine what fraction is worth keeping, necessitating complex trigger and data acquisition systems.

This poses several critical issues unique to the LHC: the read-out of all sub-detector components within strict timing constraints, the reconstruction of physically motivated particle candidates and the storage of detector information from events of interest. 
\newline

The trigger systems of the large LHC experiments play a pivotal role in their overall physics programme by dictating which events to keep or discard based only on the detector signatures of particles produced in the final state of inelastic proton-proton collisions. The physics prospects of the LHC experiments will be limited by the latency and bandwidth constraints of their trigger systems.  
Whilst machine learning (ML) has a long history of usage in the field of high energy physics (HEP), recent advances in computing have unlocked previously infeasible applications in resource constrained, real-time environments. 
Tasks such as physics object reconstruction or detector alignment and calibration can now be performed by modern ML models that are capable of interpreting the complex and high-dimensional detector data within strict latency requirements. 
\newline

The collaboration between academia and industry on the development of new tools dedicated to real-time ML applications has been crucial in improving their viability for the extreme conditions at the LHC.  Moreover, many of the tools developed for the LHC triggers can also be used in corresponding real-time regimes in industry. In a number of cases, including LHC triggers, dedicated hardware is the only way to achieve the required inference latency. 
\newline

The structure of the document is as follows: Section 2 gives a brief overview of the four large LHC experiments and the structure of their detectors. Section 3 defines the real-time environment in the context of collider experiments and discusses the motivation and challenges associated with deploying ML there. Section 4 focuses on a number of representative examples of ML applications operating in real-time at the different large LHC experiments. Section 5 introduces commonalities between the high energy physics field and industry and presents examples of cross-over or collaboration. 
\section{The Large LHC Experiments}

The detectors of the four large LHC experiments are constructed with the goal of identifying final state particles whilst maintaining sensitivity to unknown particles not predicted by the Standard Model (SM). Many of the diverse technologies and complex instruments used by the experiments have been updated in the preceeding long shutdown periods. In this section we briefly introduce the sub-detectors that make up each of the LHC experiments.

\subsection{The ALICE Detector}
The primary goal of the ALICE experiment~\cite{collaboration2008alice} at the LHC is to investigate ultrarelativistic heavy-ion collisions amongst others to to better measure the properties of the quark–gluon plasma (QGP), a deconfined state of quarks and gluons theorised to exist in the early universe~\cite{alice2014performance}. 
The ALICE physics programme is centred on the study of lead nucleus collisions (Pb-Pb), proton-lead (p-Pb) collisions as well as proton-proton interactions.

The ALICE detector is composed of a number of detector subsystems \cite{alice2019real}. The central barrel surrounding the central interaction point consists of a large solenoid magnet that generates a uniform magnetic field of up to $0.5$ T along the beam direction. Inside the magnet, a set of detectors is located surrounding the beam axis radially with dedicated detectors used for tracking and particle identification. Immediately surrounding the beam pipe is the high-precision silicon tracker (ITS). The main tracking device is the Time Projection Chamber (TPC). The Transition Radiation
Detector (TRD) and Time of Flight (TOF) subsystems provide accurate particle identification. Additional detectors, such as the muon spectrometer, are located outside the central barrel in the forward beam direction.

Ahead of LHC Run 3 the ALICE trigger system was upgraded to a streaming readout system where collision data is continuously processed and recorded~\cite{Kvapil_2773261}. The Central Trigger System (CTS) handles the data flow and synchronisation of sub-detectors before sending aggregated event information to the software-based High-Level Trigger (HLT). The HLT reconstructs the collision events, calibrates and aligns the different sub-detectors and performs data reduction.

\subsection{The ATLAS Detector}

The ATLAS detector \cite{aad2024atlas,aad2008atlas} covers nearly the entire solid angle around the collision point.
It consists of an inner tracking detector surrounded by a superconducting solenoid, electromagnetic and hadronic calorimeters, and a muon spectrometer incorporating three large superconducting air-core toroidal magnets. The Inner Detector (ID) system is immersed in a $2$~T axial magnetic field and provides charged-particle tracking in the pseudorapidity range $|\eta| < 2.5$. The high-granularity silicon Pixel detector (PIX) covers the vertex region and typically provides four measurements per track, the first hit normally being in the insertable B-layer (IBL). It is followed by the Semiconductor Tracker (SCT) and the Transition Radiation Tracker (TRT).

The ATLAS calorimeter system covers the pseudorapidity range $|\eta| < 4.9$. Within the region $|\eta| < 3.2$, electromagnetic calorimetry is provided by barrel and endcap high-granularity lead/liquid-argon (LAr) calorimeters. Hadronic calorimetry is provided by the steel/scintillator-tile calorimeter, segmented into three barrel structures within $|\eta| < 1.7$, and two copper/LAr hadronic endcap calorimeters. The solid angle coverage is completed with forward copper/LAr and tungsten/LAr calorimeter modules optimised for electromagnetic and hadronic energy measurements, respectively.
The muon spectrometer (MS) comprises separate trigger and high-precision tracking chambers measuring the deflection of muons in a magnetic field generated by the superconducting air-core toroidal magnets.

\subsection{The CMS Detector}
The central feature of the CMS apparatus is a superconducting solenoid of $6$\,m internal diameter, providing a magnetic field of $3.8$\,T \cite{hayrapetyan2023development,cms2008cms}. Within the solenoid volume are a silicon pixel and strip tracker, a lead tungstate crystal electromagnetic calorimeter (ECAL), and a brass and scintillator hadron calorimeter (HCAL), each composed of a barrel and two endcap sections. Forward calorimeters extend the pseudorapidity coverage provided by the barrel and endcap detectors. Muons are measured in gas-ionisation detectors embedded in the steel flux-return yoke outside the solenoid. At the start of 2017, a new pixel detector was installed \cite{tracker2020cms} to provide four-hit pixel coverage in the pseudorapidity range $|\eta| < 2.5$.

\subsection{The LHCb Detector}

The LHCb experiment \cite{alves2008lhcb} is a dedicated heavy-flavour experiment designed to study the properties of $b$- and $c$-hadron decays including those involving CP and flavour violation. Since its inception, however, the LHCb physics programme has grown substantially to include a large and diverse set of measurements, including studies of electroweak physics and searches for dark matter candidates.

The LHCb detector is a single-arm forward spectrometer instrumented to achieve acceptance in the pseudorapidity range, $2 < \eta < 5$ described in detail in~\cite{lhcb2015lhcb}. 
The LHCb tracking system consists of the vertex detector (VELO), the upstream tracker (UT) and the scintillating fibre tracker (SciFi). Placed immediately after the VELO and the SciFi tracker are two ring imaging Cherenkov detectors (RICH). The electromagnetic (ECAL) and hadronic (HCAL) calorimeters are placed after the RICH detector. Finally, the five muon stations lie furthest from the interaction point. 
Aside from its non-hermetic geometry the LHCb detector exhibits a number of other differences with respect to the other LHC experiments. The VELO used for tracking charged particles is located only $5.1$~mm from the beam enabling excellent impact parameter resolution. 
During the ongoing LHC Run 3 the operating luminosity of the LHCb experiment is levelled to a rate of $2 \times 10^{33}\, \text{cm}^{-2}\, \text{s}^{-1}$. At this luminosity, there are approximately $6$ proton-proton interactions per bunch crossing -- a factor of five more than in LHC Runs 1 and 2~\cite{aaij2022comparison}. 

The LHCb trigger system was upgraded ahead of LHC Run 3 and has entirely removed the hardware-based trigger step and replaced it with a second software-based trigger. The upgraded software-based trigger steps, HLT1 and HLT2, process an increased event readout using a heterogeneous computing architecture based on GPUs~\cite{aaij2020allen}.

\section{Real-Time Analysis at Collider Experiments}
\label{sec:rta_at_collider}

\subsection{The Real-Time Environment}

The data analysis workflows of the four LHC experiments generally adhere to two-stage  processing pipelines. First, events are processed synchronously, or \textit{online}, within each experiment's dedicated trigger system before they are stored. Then, the second data processing step occurs asynchronously, or \textit{offline}, across the Worldwide LHC Computing Grid (WLCG) \cite{grid2006worldwide,molina2014operating}.
The reconstruction, selection, calibration and monitoring performed online can be considered to occur in “real-time”, where power, latency and radiation constraints determine the practicability of different algorithms. Figure \ref{fig:3a} shows a simplified representation of this pipeline.

\begin{figure}[]
    \centering
    \includegraphics[height=4.5cm]{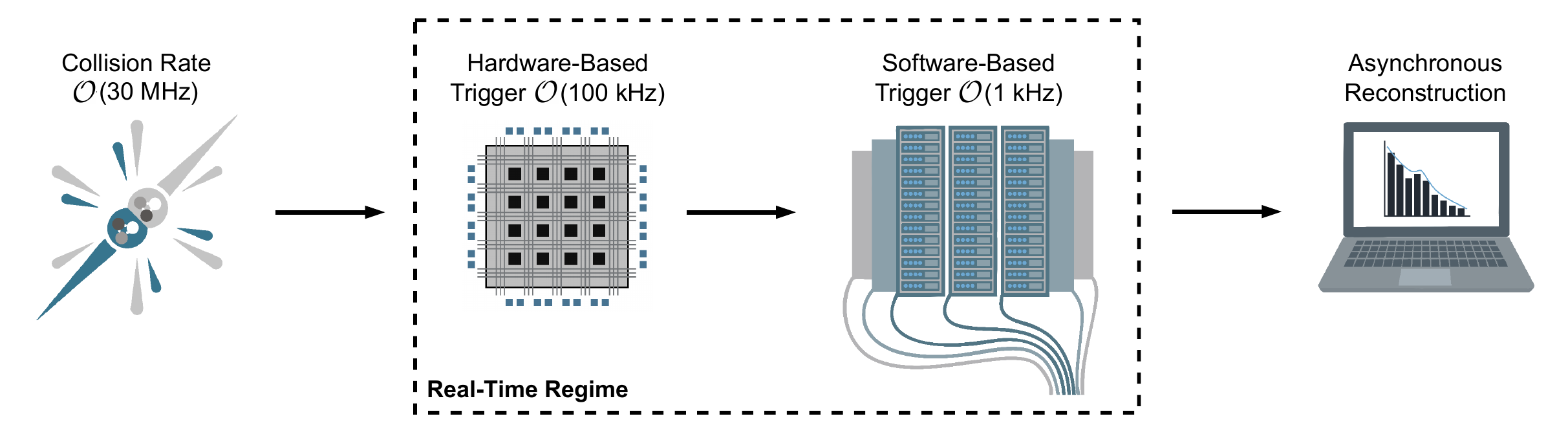}
     \caption{Simplified schematic representation of typical LHC data processing pipeline with a two-tier trigger and data acquisition system. Output event rates at each stage are typical of nominal performance of the ATLAS and CMS experiments in LHC Run 3.}
     \label{fig:3a}
\end{figure} \noindent

The development and performance of the triggers employed by the different LHC experiments is beyond the scope of this review paper, but can be found in~\cite{alice_letter, atlas_trigger, cms_l1t, cms_hlt, lhcb_upgrade} as well as in a high-level comparison in ~\cite{whitepaper_trigger,smith2016triggering}.
\newline

In the real-time regime, specifically the hardware-based trigger (the first of the two stages), it is typically not possible to read out the data from the entire detector for every event at the $30$\,MHz collision rate. Instead, it is necessary to only read-out information in selected regions of interest (RoIs) or sub-detectors that meet the strict latency requirements. The consequence of the limited and coarse information at the trigger level is a loss in the accuracy or resolution of the online algorithms which rely on lower-quality, truncated or incomplete inputs. 
In the ATLAS and CMS experiments the result of the hardware-based trigger is the rejection of approximately $99.75\%$ of the collisions, those that are accepted proceed to the second, software-based, stage of the trigger. 

In the software-based trigger, more complex methods and algorithms can be deployed. With a reduced throughput, the latency requirements are relaxed with respect to the hardware-based trigger. 
At this stage, it is also possible to read out information from all sub-detector systems and in their full granularity. Typically, the software trigger will reconstruct the physics objects that have already been estimated by the hardware trigger using more refined methods, including ML, utilising the information from the entire detector. Final trigger decisions are then made using these physics objects.
\newline

In the triggers of the LHC experiments the online reconstruction performance aims to match the offline performance as closely as possible (in some cases even being used offline directly).Often, this is not possible due to the limitations imposed by the immense volume of data produced by the LHC and the reduced fidelity of the detector readout in real-time in comparison to offline. However, due to increases in CPU performance, algorithmic efficiency and the advancement of ML models the online reconstruction performance has greatly improved in the most recent data taking period. As a consequence, the performance of online reconstruction algorithms now frequently approaches that of their offline counterparts.
\newline 

The usage of ML at collider experiments has a long history \cite{denby1988neural,aleph1999determination,collaboration1994measurement,lonnblad1991using,roe2005boosted} from simple multivariate techniques to more complex models. The use of ML has become more widespread in recent years due to the applicability of ML models in the “big data” domain at the LHC where copious collision events and Monte Carlo (MC) simulations provide immense amounts of training data. The use cases for ML in HEP have proliferated beyond enhancing signal identification and now include physics reconstruction, calibration, simulation generation, detector alignment, anomaly detection and data compression.

\subsection{Integration of ML into Existing Workflows}

The design of the LHC trigger systems undergoes near constant development. In recent years, considerable advancements have been made in the capabilities of online data processing expanding the potential frontier of real-time data processing. In order to maximise the physics potential of the prodigious dataset produced in LHC Run 3, and beyond, the online reconstruction must handle the uniquely challenging data processing conditions. This necessitates the continuous development of new reconstruction techniques or the adaptation of existing offline algorithms to comply with the requirements of online deployment.
This cyclical developmental cycle has proven to be effective in bringing offline-like performance to the real-time regime at the large LHC experiments. The progress in online performance achieved in recent years has been ameliorated by the emergence of more sophisticated ML models and dedicated hardware accelerators.
\newline 

The potential applications of ML methods within real-time data processing pipelines are diverse, spanning from relatively simple models deployed on dedicated hardware to more complex physics object reconstruction with modern neural network (NN) architectures. Furthermore, ML models can offer quick approximations to more complex, performant algorithms. 
The precise placement of such models in the dataflow is dependent on the different computational constraints and detector read-out that is possible within the wider trigger sequence.

\subsubsection{Machine Learning on Dedicated Accelerators}

An important building block of the hardware triggers used at the LHC experiments is a class of integrated circuits called field programmable gate arrays (FPGAs) that are used to perform low-level operations using logic gates, flip-flops and look up tables. Unlike application-specific integrated circuits (ASICs), FPGAs can be reconfigured after their initial manufacturing to perform modified or entirely new functions. FPGAs have been used extensively at the LHC experiments, particularly in the hardware-based triggers, where specialised hardware has been required to meet the throughput and latency specifications.
\newline

In recent years, FPGAs have evolved from specialised processors to more general hardware accelerators. One major benefactor has been the ML community - a number of tools have been developed to port ML models onto individual FPGAs. Crucially, the FPGA is able to parallelise many of the operations used inside ML models including large matrix multiplications. This has enabled significant growth in the deployment of more complex ML models onto FPGAs in the context of HEP triggers. The interplay between industry and academia has produced a number of open-source projects for deploying ML models on dedicated hardware including Conifer~\cite{summers2020fast}, FINN~\cite{umuroglu2017finn}, FwXMachina~\cite{hong2104nanosecond}, HLS4ML~\cite{fastml_hls4ml,fahim2021hls4ml} as well as proprietary platforms developed by vendors providing tools for specific devices such as Vitis AI~\cite{athur2025out} and Intel AI Suite~\cite{ahmad2022intel}.

These tools allow developers to interact with the FPGA board at an abstract level above hardware description languages (HDLs) and act as a translator between software programming languages, and low-level HDL including Verilog and VHSIC hardware description language (VHDL). These developments have removed one potential barrier to ML deployment and therefore enable a faster development cycle.

Associated developments in the training of ML models before deployment to FPGAs have seen vast improvements in model performance in resource constrained environments.
Examples of these developments include quantisation-aware training (QAT) \cite{nagel2021white}, knowledge distillation \cite{gou2021knowledge,cho2019efficacy,hinton2015distilling} and pruning \cite{cheng2024survey}. These techniques allow models to be trimmed, evaluated at lower numerical precision or to approximate the performance of larger models whilst incurring only a small degradation in performance \cite{kuzmin2023pruning,aarrestad2021fast}.
The available tools have become more widespread and are well-supported by the community at large. A number of frameworks including qKeras~\cite{coelho2021automatic,wang2021enabling}, Brevitas~\cite{brevitas} and ONNX Runtime~\cite{onnxruntime} all facilitate QAT or the mixed-precision conversion of existing ML models.

As the suite of tools for deploying ML on hardware matures and more functionality is added their usage will increase and the capabilities of the FPGAs will improve, thus allowing ML models to be brought closer to the front-end electronics of the LHC detectors in the real-time environment of the hardware-based triggers.

\subsubsection{ML Models and Software}

The evolution and development of ML libraries and frameworks has been characterised by the interplay between large technology companies and the open-source community. These libraries essentially provide a layer of abstraction between ML developers and the linear algebraic operations necessary for particular ML architectures. 
The LHC experiments, and the wider HEP community, are both consumers of and contributors to these efforts including though not limited to Caffe~\cite{jia2014caffe}, PyTorch~\cite{paszke2019pytorch}, Scikit-Learn \cite{scikit_learn}, Tensorflow~\cite{tensorflow2015_whitepaper} and TMVA in ROOT~\cite{hoecker2007tmva}.
\newline

The proliferation of ML expertise within HEP and the large number of user-friendly libraries have reduced the barriers of entry for the LHC experiments to examine more complex ML methods and reduced the amount of duplicated work related to MLOps~\cite{sculley2015hidden}. 
\newline

The data-intensive nature of the experiments open themselves up to approaches that can be refined with increased training data. In the real-time regime, a key determinant of the impact of ML will be the integration with the existing experiment codebases. 
The trigger software of each of the LHC experiments is tailor-made for the current experimental configuration. It is noteworthy that the interface between the present trigger software stacks and evolving, external libraries continues to be crucial to the success of ML deployment.

\subsection{Associated Challenges}

The use of ML in a real-time environment provides an effective and powerful method of data analysis, in some cases alleviating the computational cost of legacy or iterative algorithms. In order to achieve the required latency constraints, it is often necessary to fit the ML architecture, including all trained weights, on-chip or within the wider trigger computing network. On specialised hardware, ML models must also accommodate other essential services including those for data input and output. As previously mentioned, the latency and throughput requirements can limit the maximum complexity of the models. In spite of this, model performance in the real-time environment cannot be guaranteed to match that of offline algorithms. There exists a trade-off between the possible latency or throughput and the accuracy of ML models deployed in resource constrained environments.
\newline

The use of ML models can often reduce the need for physically motivated engineered features, rather than relying on data-driven insights that are derived from the correlation of input features. The non-parametric nature of some ML models means that it can be difficult to interpret the model output for a given set of inputs. This can obscure the underlying physics models and processes that govern the collisions measured at the LHC and elsewhere. One recent direction of research in HEP has been in encoding domain-specific inductive biases from HEP within ML models directly. This includes enforcing a monotonic response to input features, equivariance under some transformation of the inputs~\cite{cohen2016group,spinner2024lorentz} or similar physics-informed requirements~\cite{toscano2025pinns}. 
\newline

Related to the difficulties of interpretation are the susceptibility of many ML models to out-of-distribution effects or edge cases. Deployment in the data analysis workflow at the LHC experiments can lead to unforeseen failure modes of the model due to the large statistics involved and the potential domain shift from model training to inference. Therefore, it is crucial that the model outputs are verifiable and interpretable such that unexpected behaviour can be both identified and rectified. The ability to understand the inner workings of ML models is essential both to understand the underlying physics and to strengthen the certainty with which claims can be made a posteriori. The reliability of a particular model takes on increased importance at collider experiments, where final states of inelastic collisions are governed by quantum field theory and can only be inferred statistically. The interpretability and explainability of ML models is an active area of research, particularly in the HEP field \cite{wetzel2025interpretable,arzani2024interpreting}.
\newline

Another challenge that accompanies the use of ML in the HEP domain is the inherent reliance on MC simulation. The supervised training of large ML models is predicated on the existence of large datasets with reliable “truth” labels that are used to discover the relationship between different input variables and the desired output. Collider experiments have used MC simulation for a variety of applications outside of ML training, but the advent of larger and more complex ML models and their corresponding increase in performance has meant that ML performance is now reliant on the MC generators used to produce the simulation. It is possible that sophisticated ML models are able to find or recognise relationships that exist in simulation but do not appear in real particle collisions, therefore degrading the model performance in its application to real data. This effect would manifest as a systematic deviation related to the mismodelling of the simulation. Though this problem is an open area of study, the large LHC experiments employ several strategies to mitigate it, including: validation using adjacent control regions, weakly supervised learning using real data and parameterising known nuisance parameters within the network architecture \cite{guest2018deep}. 
\newline

The challenges detailed above are exacerbated in the real-time regime due to the nature of the  trigger and data acquisition systems at the LHC experiments; errors in event selection are unrecoverable since particle collisions are either saved to storage or discarded. In this way, there is no way to compensate for errors during data taking periods - if an ML model has some inaccuracy or bias due to it's MC simulation labelled training corpus or an unexpected edge case then entire classes of collision events may be irretrievably rejected.

\section{Examples of ML for Real-Time Analysis}
\label{sec:current_and_future_uses}

\subsection{Physics Reconstruction}

\subsubsection{Track Reconstruction in LHCb}
At the LHCb experiment, the reconstruction of charged particles in the tracking system plays a crucial role in its physics programme.
Ahead of LHC Run 3, the LHCb trigger system was migrated to a fully GPU-based first-level trigger (the Allen system~\cite{aaij2020allen}) that reduces the initial dataflow throughput from $5\,\text{Tb}\,\text{s}^{-1}$ to less than $200\,\text{Gb}\,\text{s}^{-1}$~\cite{morris2024first}. The use of GPUs at this stage of the real-time data processing pipeline opens up the potential for deploying complex ML-based algorithms to aid in the reconstruction and classification of events. One example of this is the ongoing development of a Graph Neural Network (GNN) to identify tracks in the VELO sub-detector.
\newline 

In LHC Run 3, there are approximately $5$ proton-proton collisions per bunch recorded by LHCb in the VELO sub-detector acceptance. The GNN reconstruction, called ETX4VELO~\cite{correia2024graph}, implements track finding in five steps: hit graph construction, edge classification, edge graph construction, triplet classification and track construction. The different steps improve the efficacy of the ETX4VELO algorithm for tracks with challenging experimental signatures, including intersecting tracks or shared hits, skipped VELO planes, multiple hits per plane and tracks containing electron-positron pairs. Functionally, the algorithm can be broken down into several sequential models: an embedding DNN, an edge classification GNN with multiple rounds of message passing and a Weakly Connected Component (WCC) step where the final tracks are retrieved.
\newline
The performance of ETX4VELO rivals the track reconstruction used in the Allen framework across all track types, with particularly pronounced improvement for electron tracks. Moreover, the number of fake tracks produced is reduced to less than 1\%, compared to previously more than 2\%. Critically, ETX4VELO detects 15\% more electrons that only leave tracks in the VELO. 
Work is ongoing to explore the deployment of ETX4VELO on dedicated FPGA accelerators \cite{giasemis2025comparative} using the HLS4ML library. Such an extension could provide   further improvements including enhanced computational performance, greater power efficiency and higher throughput while more easily accommodating the requirements of future HEP experiments.

\subsubsection{Electron and Photon Identification in ATLAS}
In the ATLAS experiment trigger system, there are several fast algorithms employed as preselections to quickly reject dominant background sources before more complex and performant algorithms are run downstream to produce the final event selections~\cite{atlas2024atlas}. One example of an ML-based preselection is the Ringer algorithm~\cite{SpolidoroFreund_2675025}, which is used to identify electrons and photons in the ATLAS calorimeter. 
\newline

The Ringer is a DNN that takes as input the energy sum of cells in $\mathcal{O} (100)$ concentric rings centred on the cell with the highest energy in each calorimeter sampling layer. The ring inputs exploit the approximately conical structure of electromagnetic showers and their lateral development within a single calorimeter layer. Further, the concentric rings are statistically dependent both laterally and longitudinally which allows the DNN to leverage their correlations.
The concatenated vector containing the normalised energy sum of each concentric ring is fed into an ensemble of simple MLPs defined for each $E_T \times \eta$ region in the calorimeter.

The Ringer algorithm deployed in the ATLAS software trigger was marginally slower than the previous cut-based approach. Despite this, its greater rejection efficiency  resulted in a $50\%$ reduction in the overall CPU demand for single electron triggers.

\subsubsection{Tau Reconstruction in CMS}
The tau lepton, with its large mass and short lifetime, typically decays before reaching the active material of LHC experiments. It is therefore necessary to reconstruct this lepton via its decay products alone. Two-thirds of tau lepton decays result in hadrons and a single tau neutrino hence, their identification partially relies on standard jet reconstruction.
In its LHC Run 3 trigger system, the CMS experiment utilises a custom DNN incorporating convolutional layers to identify hadronically decaying tau candidates~\cite{varghese2023cms}. This method is able to simultaneously reject the quark or gluon initiated jets, electron and muon backgrounds that contaminate the hadronic tau signature. Previously, separate discriminators were used to reject jets, electrons and muons respectively.
\newline 

The DNN architecture, called DeepTau~\cite{tumasyan2022identification}, utilises a combination of high-level input features as well as local calorimeter information to discriminate the hadronic tau candidates from all relevant backgrounds at once. The DeepTau network consists of three subnetworks that each handle a different data input: the outer cells, inner cells and high-level variables. 
The subnetwork outputs are concatenated and passed through four hidden layers before yielding a final output vector that can be interpreted as the probability that the object belongs to each class $\{ \tau,\, \text{jet},\, \mu,\, e \}$. 
The subnetworks processing the inner and outer cells are based on convolutional layers that act on embedded features derived from the grid-like cell information. The convolutional layers successively reduce the grid size before it is concatenated and passed to the output dense layers.  The 47 high-level variables pass directly through 3 dense layers before concatenation. 
\newline

The DeepTau network achieves an improvement in efficiency of between $10\%$ and $30\%$ for a given jet rejection, while the efficiency of tau candidates passing the loose identification threshold increases by $14\%$ with DeepTau compared to the baseline performance.

\subsubsection{Jet Flavour Tagging in ATLAS}

The identification of jets initiated by heavy-flavour hadrons (\ \(b\)- and \(c\)-hadrons) is a fundamental component of the physics programmes of the different LHC experiments. Efficient discrimination between heavy-flavour jets and those arising solely from light quarks or gluons is essential to suppress the overwhelming QCD dijet background in proton–proton collisions.  Jets may also contain decay products of electroweak bosons (e.g.\ \(W\), \(Z\)), top quarks, \(\tau\) leptons or even the Higgs boson.  Experiments around the LHC apply algorithms to identify or “tag” these jets using their distinguishing physical properties to separate them from light-flavour hadrons or gluon-induced jets. 
\newline

In LHC Run 3 the ATLAS experiment has introduced new ML models to replace existing flavour tagging algorithms that previously relied on high-level observables engineered specifically for jet flavour classification. The new ML models utilise low-level features including parameters related to the primary vertex and the tracks associated to each jet to discriminate between the different jet flavours. The collection of low-level features implicitly contains information about the hadron lifetime and the secondary vertex mass. 
\newline

The models now used in the ATLAS trigger are based on either Deep Set (fastDIPS~\cite{atlas2023fast}) or Transformer (GN1~\cite{atlas2025configuration} and GN2~\cite{atlas2025gn2}) architectures. These permutation invariant network architectures directly combine the track and jet information to infer the jet flavour. Both architectures embed the the input features before combining them via a summation or attention mechanism respectively. The model output is a multi-class probability score across the different jet flavours that is combined into a single discriminant that is used to make a final determination. 
The fastDIPS and transformer-based models are deployed in the software-based trigger in ATLAS. In the real-time regime the model performance on jets with low transverse momenta is particularly important because of the relatively large cross-section of multijet production at the LHC. Therefore the flavour-tagging algorithms can be used to reduce the light-jet background and thus the event rate to a feasible level. 
Moreover, for some signatures these ML models can be used as an approximate preselection applied earlier in the software-based trigger to provide a preliminary reduction in CPU cost. The models can be executed using coarse or unrefined input features and still provide an acceptable level of rejection that prevents the execution of more computationally expensive methods downstream.

\subsubsection{Pile-up Mitigation}

At the luminosity at which the LHC collides bunches of protons it is likely that multiple proton-proton interactions occur concurrently in the same bunch crossing.
The additional proton-proton interactions that occur alongside the most energetic primary vertex (referred to as the hard scatter vertex) are considered ``pile-up" interactions. The average number of simultaneous collisions in the ATLAS and CMS experiments is estimated to exceed $50$ in LHC Run 3~\cite{Experiments_2903107}. 
\newline

The goal of the LHC experiments is, in general, to analyse only the hard scatter vertex from each bunch crossing, in order to most benefit from the operational centre of mass energy of the LHC. The pile-up interactions deposit energy in each of the detectors and intermix with the signals produced by the interaction of interest. Most of these pile-up interactions are soft inelastic collisions that produce relatively low-energy deposits with an isotropic distribution. In order to be able to accurately distinguish the most energetic interaction in each bunch crossing, the pile-up contributions must be suppressed. This suppression is crucial in a real-time environment where a decision is made about whether to store an event or not based on the candidate physics objects present in the event. Particles originating from pile-up interactions can erroneously allow events to pass trigger hypotheses - thus contaminating the final event selection. It is critical to the operation of LHC triggers that the additional pile-up particles are identified, if possible, and accounted for in real-time.
\newline

One approach, PUMML~\cite{komiske2017pileup}, aims to use CNNs to remove the pile-up contamination from the interaction of interest. The PUMML algorithm targets the transverse momenta of neutral particles originating from the primary vertex. Neutral particles represent a challenging signature to reconstruct as there is no corresponding track pointing back to the primary vertex. The network input is a $36\times36$ pixel image centred on a single jet axis with three channels. Each channel corresponds to a different source of transverse momentum (primary vertex charged, pile-up vertex charged and total neutral). 
When evaluated on simulation the PUMML algorithm matches the efficiency of current rule-based algorithms at removing neutral pile-up particle transverse momenta. 

Another, more recent application of ML to pile-up mitigation is PUMA~\cite{maier2022pile}, which uses a sparse transformer to model pile-up contributions. PUMA reconstructs the hard energy fraction of each reconstructed particle candidate in an event such that it may be corrected to reduce pile-up contamination. This problem formulation accommodates the finite spatial and energy resolution of the detectors and that it may not be possible to fully separate contributions from the primary or pile-up vertices. PUMA uses a sparse self-attention mechanism to reduce the time and memory complexity of full self-attention, since the number of pile-up particles can exceed the constraints of even large GPUs.
The PUMA model outperforms classical pile-up suppression methods across the board with a $20\%$ improvement in the resolution of missing transverse momentum per event.

\subsubsection{Identifying Heavy-flavour Decays in LHCb}

In LHC Run 3, the LHCb topological trigger was updated to include a monotonic Lipschitz NN \cite{delaney2024applications, schulte2023development} as a final selection criterion in the identification of displaced \(b\)-hadron decays. After an initial cut-based preselection is used to discard prompt and soft background candidates the monotonic Lipschitz NN is used to make an additional selection. Two identical models are trained separately, targeting two- and three-body decays respectively, which are defined based on the multiplicity of charged particles reconstructed in the decay products of the $b$-hadron secondary vertex. 

The networks take as input a vector of features derived from the final-state tracks of the multi-body decay products that capture the overall topology and kinematics of the system. Observables include the candidate transverse momentum, decay-vertex fit quality, flight distance,  distance of closest approach of charged tracks, multi-body corrected mass value \cite{abe1998measurement} and the impact parameter with respect to the primary vertex.
The NN enforces a constraint on the maximum value of the Lipschitz constant of the learnt decision-boundary function. The bounding of the Lipschitz constant ensures that events that are in close proximity in the feature input space are also close in the final decision space of the network. By setting a strict upper limit on the classifier response variation, the magnitude of the effects of finite detector resolution and other experimental instabilities are limited.

The monotonic Lipschitz architecture offers improved robustness under varying detector conditions and contains physically motivated inductive biases. The performance of the ML model is improved with respect to the previously employed BDTs used in LHC Run 2, whilst also satisfying the increased bandwidth and latency requirements of the improved LHCb trigger in LHC Run 3.

\subsection{Object Calibration}

\subsubsection{TPC Calibration in ALICE}

Before LHC Run 3 the ALICE trigger and data acquisition system was upgraded to a streaming readout - data is continuously processed and recorded for every collision~\cite{hellbar2021reconstruction}. This novel readout system represents a paradigm shift in data management at an LHC experiment and presents a number of challenges and opportunities for its future operation. Among the challenges, the real-time calibration of detector signatures will be of paramount importance in order to ensure optimal detector operation and rapid feedback in case of detector malfunction. 
\newline

The TPC is the main detector for tracking and particle identification in ALICE. The TPC identifies charged particles directly by reconstructing their trajectories as they pass through the TPC volume. The signals are amplified by Gas Electron Multipliers (GEMs). The space-charge density within the TPC depends on the number of pile-up ions in one ion drift time. The expected distribution of space-charge density at 50 kHz of lead-lead interactions can be derived, investigated and tested using MC simulations.
 
The calibration of the TPC aims to correct the average space-charge distortions and the distortion fluctuations during both synchronous and asynchronous reconstruction. The timing constraints for the distortion fluctuation calibrations during readout ($\mathcal{O}$ (10\,ms)) can be met whilst using a fast ML model. A CNN model has been developed to predict the space-charge distortion fluctuation corrections~\cite{gorbunov2021deep} based on a U-net architecture \cite{ronneberger2015u}, where the input image is a $3$-dimensional representation of the space-charge distortions fluctuations map in a $90\times 17\times 17$ or $180\times 33\times 33$ grid-like representation. 
Development of the CNN is ongoing though initial performance appears to depend on the distance from the fiducial boundaries of the TPC.

\subsection{Anomaly Detection}

Anomaly detection (AD) in HEP can be described as the process of identifying out-of-distribution events that broadly exhibit deviations from the SM with no a priori assumption on the specific alternative hypothesis. This includes both the detection of erroneous events affected by stochastic noise inherent to the LHC subdetectors as well as rare events, potentially containing physics beyond the standard model (BSM).
\newline

These two branches of AD, while correlated, address different classification problems. The first classifies events that have been impaired as a result of noise or other undesirable detector effects, ideally at individual component levels. To maintain a high data taking efficiency and prevent the contamination of datasets, these events must be monitored and, if necessary, removed. This process is broadly referred to as Data Quality Monitoring (DQM). In HEP, noisy events are considered anomalous and standard detector performance is presumed to make up the bulk of the in-distribution events. The identification of these anomalous noisy events not only aids in cleaning the datasets collected by the detector, but also helps in understanding novel detector effects, such as noise bursts or high-voltage trips that can occur during data taking.

The second branch involves the identification of rare and unusual events, and is used as a tool in new physics searches by isolating events that exhibit behaviour inconsistent with the SM. These events can be saved for further analysis whilst remaining agnostic to the specific type of BSM physics model~\cite{belis2024machine}.

The operation of these AD methods occurs both online and offline in HEP experiments. However, the use and performance of AD models in real-time has proliferated in recent years as model capabilities and inference speed have improved. This has enabled more AD models to be deployed earlier in the event selection process at the LHC experiments, enabling less biased selections in model-agnostic searches for new physics and greater coverage with DQM earlier in the trigger system.

\subsubsection{Data Quality Monitoring in ATLAS and CMS}
\label{sec: dqm}
The rate of collisions at the LHC during the Run 3 data taking can be up to $30$\,MHz, this represents an enormous flow of data that must be filtered, compressed and stored. Monitoring the data quality is therefore an essential task. The current DQM system in most LHC experiments is split into two categories: during data-taking (online) and asynchronously after data storage (offline). Most often both online and offline monitoring involve the comparison of histograms containing event-level variables to a pre-defined reference by a human expert. In the context of RTA, the online DQM is often dependent on a fraction of representative events written out to dedicated calibration or monitoring data streams that are promptly reconstructed and used to more quickly identify issues that could immediately affect data-taking efficiency. 
\newline

The ATLAS and CMS experiments have both begun to examine the effectiveness of unsupervised ML models to assist human experts within DQM systems~\cite{pol2019detector,brinkerhoff2025anomaly}. A common example is the use of autoencoder (AE) architectures that are trained on examples of nominal detector performance to reconstruct the original input and therefore do not need truth labels.
AE models typically consist of two neural networks used in sequence to first ‘encode’ the input to some lower-dimension latent space and then to ‘decode’ the resultant latent vector representation back to the original input space. The performance of the AE compression-reconstruction is evaluated using a reconstruction loss which compares the model output to the original input.
\newline

In CMS a CNN-based AE model focussing on the ECAL was deployed online in the DQM workflow during 2022 in LHC Run 3~\cite{cms2024autoencoder}. 
Occupancy histograms from the barrel and endcap sub-detectors are processed as two-dimensional images and a residual neural network (ResNet)~\cite{he2016deep} is used to first compress the image into an informative low-dimensional representation before the decoder upsamples the latent representation back to the original image size. The mean squared error between the input and the AE reconstructed output is used to optimise the model’s performance on standard detector operation, as defined by experts.
The addition of this AD model into the DQM has proven to be highly efficient with anomaly identification estimated to be approximately $99\%$ in validation datasets with controlled noise injections. The addition of the CNN-based AE has complemented the existing DQM by simplifying decisions taken by detector experts and reducing the number of reported false alarms. 
\newline

The ATLAS experiment has developed a different AE-based model for DQM~\cite{ATL_DAPR_PUB_2024_002}, based on recurrent neural network (RNN) blocks called long-short term memory ~\cite{Hochreiter97} (LSTM) for both the encoding and decoding of the detector information.  

In this application, data from the LAr calorimeters in the ATLAS detector are treated as multidimensional time-series where periods of nominal detector performance are punctuated by scarce, brief intervals of anomalous activity due to stochastic noise. The majority of noise bursts in the LAr calorimeter have a duration of less than $60$\,s.

The LSTM-based AE model classifies a time-series sample as standard detector performance or noisy in an unsupervised regime. The LSTM-based model is able to capture long-term dependencies and correlations within the data, in the context of DQM this is pertinent since the noise bursts in the LAr calorimeter corrupt data over contiguous periods of data-taking. 
The input data are comprised of normalised features related to clusters of calorimeter cells taken from different regions of the detector. 
The preliminary results from this work are promising and require few modifications to be run in real-time in the ATLAS trigger. Moreover, the performance was observed to extend to heavy ion collisions without retraining.

\subsubsection{New Physics Searches in CMS}
\label{sec: newPhysAD}
A fundamental assumption of AD in the context of searches for new physics is that any physics signature produced in the LHC detectors will be distinct from the overwhelming SM background present. Traditional physics analyses conducted at the large LHC experiments typically depend on particular signatures that arise from a predetermined set of signal hypotheses that can be simulated. AD-based BSM searches on the other hand intend to exploit the potential dissimilarities between new physics and SM processes to evaluate any BSM model in the relevant phase space. In this way, AD-based new physics searches can be used as a safeguard against biasing the direction of searches in areas with theoretical motivation, thereby allowing under-investigated or unposed models to be probed further.

In addition, AD-based physics searches have benefited from the increased inference speed of ML models that allow AD scores to be calculated at the trigger-level. Therefore allowing the experiments to cast a wider net with reduced dependence on the standard set of triggers. The triggers in LHC experiments have been carefully optimised to increase both the sensitivity to their existing search programmes and the precision of their SM measurements. Deploying AD algorithms earlier in this chain allows more events up to the maximum collision rate of $30$\~MHz, to be examined before they are discarded by the model-dependent trigger and data acquisition algorithms.

\begin{figure}[]
    \centering
    \includegraphics[width=7.5cm]{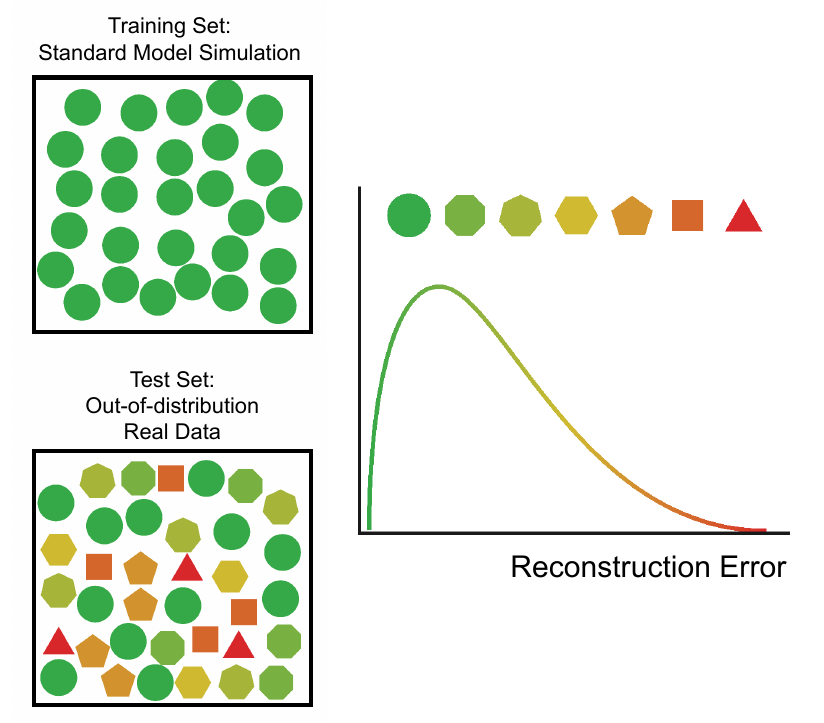}
     \caption{Illustration of unsupervised learning in anomaly detection training for HEP. Out-of-distribution samples result in a large reconstruction error at inference. Often, simulation is used to define the in-distribution training data, which is populated by standard model physical processes.}
     \label{fig:4a}
\end{figure} \noindent

The CMS experiment has developed and deployed two complementary AD networks to be placed in the initial, hardware-based trigger that is composed of custom hardware with a strict latency requirement of $4$~$\mu$s~\cite{gandrakota2024real}. 
CICADA~\cite{CMS_DP_2023_086} (Calorimeter Image Convolutional Anomaly Detection Algorithm) is a two-dimensional CNN that translates calorimeter region energy deposits into pixels in a $18\times14$ image. The training data is made up of zero-bias data events and since the network does not generalise well, by design, events that do not resemble the standard model training set will have a higher reconstruction error. 
AXOL1TL~\cite{CMS_DP_2023_079} (Anomaly eXtraction Online Level-1 Trigger aLgorithm) is another ML model used by CMS to separate rare and exotic events in real-time. The AXOL1TL architecture is a Variational Autoencoder (VAE) that is also trained on zero-bias data and takes up to 19 coarsely-defined physics objects reconstructed in the L1 trigger as input. The loss term for the VAE includes a full regularisation term (KL divergence) as well as the MSE loss. %
Both models were trained using QAT and synthesised using the HLS4ML library discussed in Section 3. Both algorithms were then each deployed on a single Xilinx Virtex-7 FPGA board. The CICADA model has an approximate latency of $\mathcal{O}(100)$\,ns whereas for AXOL1TL the inference speed is just $50$\,ns.

The novel use of these models in the trigger demonstrates that model-independent methods for AD are possible in real-time applications, even in environments that are severely resource constrained and have extreme latency requirements.

\section{Synergies between HEP and Industry}
\label{sec:industry}

\subsection{Common Challenges of Real-Time Analysis}

Even in the era of big data, the computing challenges experienced at the LHC are unique. Proton-proton collisions occur with a frequency of up to $30$\,MHz in the LHC, corresponding to a streaming bandwidth of approximately $1$\,PB/s. This rate of data generation is among the largest in the world and, coupled with the imposed latency requirements, results in computing challenges without precedent. The trigger systems used in LHC experiments are subject to some of the largest global data throughputs whilst enforcing some of the shortest latency requirements. 
\newline

As previously discussed, it is not possible for LHC experiments to record all events due to both throughput and storage limitations. Moreover, the overwhelming majority of LHC collisions produce well-understood or uninteresting final state particles — it would not be worthwhile for LHC experiments to record these events at all. To accommodate the constraints of the readout and storage systems and advance the physics programme of the experiments, the trigger accepts on average only 1 in 30,000 collision events. 

There are many examples from industry that also confront strict data throughput and latency requirements. Machine learning has become one of the most prominent tools used in real-time applications in industry in a diverse set of domains, including, though not limited to meteorological prediction and simulation, medical imaging, cybersecurity monitoring and natural language processing. 

Industrial applications, similarly to the LHC, contend with a huge volume of data and must synthesise the vast amounts of information into a more manageable size in order to maximise the benefits of the scale of available data. The shared challenges between the HEP field and industry engender shared solutions to common problems, such as open source software frameworks, accessible tools for high-performance computing and cross-platform support among vendors of hardware accelerators.

\subsection{Applications in Industry: Time-Series Anomaly Detection}

In industry, one real-time application of ML models is the detection of anomalies in time-series data. For example, IBM France is pursuing the development of time-series AD methods for fraud detection among financial transactions. In this application the out-of-distribution “anomalies” are the fraudulent transactions that should be identified and flagged.
The financial transaction data is processed in a time-series, preserving the auto-correlations between the data. Moreover, the temporal dependence of fraudulent transactions means that unusual patterns develop over time, allowing ML models to capture and exploit their temporal relationship.

There are several challenging aspects to the development of such an AD model for example that, in a similar way to the DQM AD in HEP detectors, anomalies, or cases of fraud, are so infrequent that there is a critical lack of statistics of anomalous events. This rarity and class imbalance can make it difficult to assess the performance of AD models. In the financial industry, less than $1\%$ of credit card transactions are fraudulent, but they can incur huge financial losses on the order of billions of euros \cite{leborgne2022fraud,bank_sixth_2020}.
Data in the financial industry, as in many other domains, are protected by strict confidentiality and privacy regulations constraining the size and completeness of datasets. This, alongside the fact that it is very difficult to label data, means that trustworthy labelled data are scarce.
\newline

IBM Research France takes a two-fold approach to tackle these challenges, by actively investigating the use of unsupervised models for AD and producing auxiliary synthetic financial transaction datasets.
The development of an unsupervised AE-based fraud detection model utilises the modern transformer architecture to encode sequential time series data before a decoder calculates the reconstruction error with respect to the initial input. The AE with iTransformer~\cite{liu2023itransformer} encoder compares favourably to alternative established classical and NN-based AD methods including MERLIN~\cite{nakamura2020merlin}, Isolation Forest~\cite{liu2008isolation}, fully connected NNs and LSTMs. The performance of transformer-based architectures on numerical time-series data is still an active area of research, particularly the ability of such models to capture long-range dependencies. The interpretation of the reconstruction errors as anomaly scores plays a crucial role in model performance in this setting, various metrics, anomaly thresholds and evaluation protocols (when a fraudulent warning is triggered) are considered.

The production of synthetic financial data is pursued to mitigate the unreliability and lack of labelled data. Unlike HEP, the financial industry does not always have large, open datasets with accurate ground truth labels, it can be necessary for a human expert to manually generate such labels whereas the HEP field relies on advanced simulations underpinned by theoretic predictions of known particle physics phenomena. The application of ML in the financial industry could ameliorate the lack of reliable training statistics, one example is the generation of synthetic data for fraud detection. In a first step, a bank fraud simulation system is built using Markov chains designed to model both legitimate and fraudulent user behaviours within a financial environment. Then, the simulation system leverages Large Language Models (LLMs) in a multi-step generation following nominal and fraudulent strategies with respect of temporal and other dimensions of the described activities captured in the dataset. Whilst this work remains ongoing it demonstrates promising initial results \cite{feillet_genovae_2025}.

\subsection{Applications in Industry: Computer Vision}
One prominent use of ML models in recent years has been computer vision (CV); the remarkable efficiency with which NNs can extract and distill high-level information from images and point clouds means that there is a plethora of real-world applications. Many of these applications, like HEP detectors, have strict latency, bandwidth or memory requirements that impose constraints on the deployment of ML models.

The HLS4ML package, as mentioned in Section 3, can be used to facilitate the deployment of  trained ML models on FPGAs targeting low latency and low power environments. The impact of the project has been felt beyond the wider scientific community and is being used in a variety of different industrial settings, in this section we present two examples.
\newline

A joint research project conducted by CERN and Zenseact, explored the use of FPGA hardware accelerators for the real-time semantic segmentation of images from camera sensors on-board autonomous vehicles~\cite{ghielmetti2022real}. The recent advancements in modern deep learning CV applications can improve the safety and efficiency of autonomous driving by allowing the vehicle to analyse and respond to its external environment more accurately and faster. 

Using the HLS4ML package, FPGAs were evaluated as a hardware benchmark for CV models based on simple CNN architectures~\cite{aarrestad2021fast}. The ENet architecture~\cite{paszke2016enet} was used as a baseline during QAT to attain a suitable compression of the model. An additional constraint limits the total number of filter kernels permitted in the model, similarly to filter ablation, where an optimum value was found by balancing performance and resource requirements. 

The inference time of the updated ENet model was tested on a single Zynq Ultrascale+ MPSoC device using a ZCU102 development kit and found to be less than $5$\,ms per image, faster than previous FPGA implementations. The best performing model was found to use less than $30\%$ of the total resources of the FPGA.
\newline

Another example of a real-time CV application is the EU project Edge SpAIce that aims to monitor the plastic pollution in oceans using a satellite-mounted camera and ML techniques. Earth observation satellites are capable of transmitting data back to Earth with a limited bandwidth and power usage. Similarly to the resource constraints faced in HEP detector readout systems, FPGAs can be used to alleviate these issues. Edge SpAIce uses fast NN inference aboard the satellite to segment the images into different object classes such that only images relevant to a particular class are transmitted back to Earth~\cite{tzelepis_2024_13865939}. This classification represents an initial type of bandwidth reduction before data transmission. 

As before, the HLS4ML package was used to move the U-net architecture based on CNN layers onto an FPGA hardware accelerator. The task is formulated such that the satellite images ($256 \times 256 \times 3$) are segmented to identify pixels covered by clouds. The supervised training, using per-pixel truth labels, proceeds in a quantisation-aware fashion and uses knowledge distillation from a larger, teacher model to reduce the number of learnable parameters in the smaller, student model. Using the HLS4ML workflow results in a performance improvement of a factor greater than $8$ with respect to the Vitis AI tool whilst adhering to the resource constraints of the ZCU102 development kit.

\section{Conclusion}
The trigger and data acquisition systems at the large LHC experiments have undergone significant changes since LHC Run 1. The capabilities and efficacy of the event selection pipeline have been considerably enhanced by the use of ML techniques. The level of performance achieved in the real-time regime has markedly improved and in many cases approaches the state-of-the-art achieved offline.

Collider experiments and the HEP field more generally have long utilised ML models to ameliorate signal identification. Indeed, with the prodigious datasets provided by MC simulation and real collisions, the LHC has ample training data to exploit and train ML models. However, it is only recently that the trigger systems and real-time reconstruction have been able to leverage these advancements. 
The challenges of deploying complex algorithms in real-time environments had resisted the progress made for algorithms employed during asynchronous reconstruction.

In spite of the constraints imposed in real-time environments, technological and algorithmic advancements have allowed LHC experiments to deploy ML within their trigger systems. This trend looks set to continue as more applications and opportunities arise. This work presents examples of the usage of ML models in real-time data taking environments across the large LHC experiments.
Beyond improving the performance of current algorithms used in the trigger ML in the real-time regime may also extend the capacity of the LHC experiments’ physics programmes by offering model-agnostic methods for anomaly detection. Enlarging the phase-space that can be searched by relying on events that exhibit discrepancies with respect to the SM could inform future searches for new physics.

The use of ML in real time is not limited to the HEP field and there are may industrial applications that necessitate similar strategies. The collaboration between CERN and industry partners arising from the unique demands of trigger and data acquisition systems at the LHC have propelled the community forward by providing open, accessible and well supported libraries for hardware deployment of ML. Continuing knowledge exchanges between research and industry will be crucial in order to continue this progress for LHC triggers as well as real-time industrial applications.

\clearpage
\section*{Acknowledgements}
This work is part of the SMARTHEP network and it is funded by the European Union’s Horizon 2020 research and innovation programme, call H2020-MSCA-ITN-2020, under Grant Agreement n. 956086. 
We gratefully acknowledge Teng Jian Khoo, Maurizio Pierini and Chris Scheulen for their comments and advice for this whitepaper. We also thank the wider participants of the SMARTHEP network for their insight and feedback.

\bibliographystyle{JHEP}
\bibliography{references.bib}

\end{document}